# Two-Dimensional Material Nanophotonics

Fengnian Xia[1*], Han Wang[2], Di Xiao[3], Madan Dubey[4], and Ashwin Ramasubramaniam[5]

[1]*Department of Electrical Engineering, Yale University*
*15 Prospect Street, New Haven, CT 06511*
[2]*Ming Hsieh Department of Electrical Engineering, University of Southern California*
*3737 Watt Way, Los Angeles, CA 90089*
[3]*Department of Physics, Carnegie Mellon University*
*5000 Forbes Avenue, Pittsburgh, PA 15213*
[4]*Army Research Laboratory, Sensors and Electron Devices Directorate*
*2800 Power Mill Road, Adelphi, MD 20783*
[5]*Department of Mechanical and Industrial Engineering, University of Massachusetts, Amherst*
*160 Governors Drive, Amherst, MA 01003*

***fengnian.xia@yale.edu**

Abstract. **The emerging two-dimensional (2D) materials exhibit a wide range of electronic properties, ranging from insulating hexagonal boron nitride (hBN), semiconducting transition metal dichalcogenides (TMDCs) such as molybdenum disulfide ($MoS_2$) and tungsten diselenide ($WSe_2$), to semi-metallic graphene. The plethora of 2D materials together with their heterostructures, which are free of the traditional "lattice mismatch" issue, brings new opportunities for exploring novel optical phenomena. In this review, we first discuss the optical properties and applications of a variety of 2D materials, followed by two different approaches to enhance their interactions with light: through their integration with external photonic structures and through their intrinsic polaritonic resonances. Finally, we cover a narrow bandgap layered material, black phosphorus, which serendipitously bridges the zero gap graphene and the relatively large-bandgap TMDCs such as $MoS_2$ and $WSe_2$. The combination of these materials and the approaches for enhancing light-matter interaction offers the promise of scientific discoveries and nanophotonics technologies across a wide range of electromagnetic spectrum.**

# I. Unique properties of 2D materials

Many layered materials in their bulk forms have been widely known and utilized for a long time. For example, graphite and molybdenum disulfide ($MoS_2$) are used as dry lubricants due to their layered nature: atoms are strongly bonded within the same plane but weakly attached to sheets above and below by van der Waals force. This weak interlayer interaction makes the extraction of single or few-layer of atoms possible, leading to the burgeoning research on 2D materials[1,2,3,4,5,6,7,8,9,10,11,12,13,14,15,16,17,18,19,20,21,22] including their nanophotonic properties and applications. In this review, 2D materials are broadly defined to include multilayers, heterostructures, and layered thin films whose total thicknesses vary from an atomic layer to tens of nanometers. "Two-dimensional (2D)" here implies that the thickness of the materials under investigation is orders of magnitude smaller than the wavelength of the light involved.

Compared to traditional three-dimensional photonic materials such as gallium arsenide (GaAs) and silicon (Si), 2D materials exhibit many promising properties: (1) their surfaces are naturally passivated without any dangling bonds, making the integration of 2D materials with photonic structures such as waveguides[23,24,25,26] and cavities[27,28,29,30,31] easy. It is also possible to construct vertical heterostructures using different 2D materials without the conventional "lattice mismatch" issue, since layers with different lattice constants in heterostructures are only weakly bonded by van der Waals force as in layered bulk materials. (2) Despite being atomically thin, many 2D materials interact with light strongly. For example, a single layer of molybdenum disulfide ($MoS_2$) absorbs around 10% at excitonic resonances (615 and 660 nm)[32]. (3) 2D materials can cover a very wide electromagnetic spectrum due to their diverse electronic properties, as shown in Figure 1a.

The gapless graphene interacts with light from microwave to ultraviolet[33,34], making it a potential candidate for various light detection, modulation, and manipulation functions. However, its metallic nature prevents the realization of efficient light emitting devices using graphene. On the contrary, single layer transition metal dichalcogenides such as $MoS_2$ and $WSe_2$ are direct bandgap semiconductors[14,15] as shown in Figure 1c. They exhibit encouraging light emitting properties primarily in near-infrared wavelength range dominated by excitons and trions, due to the strong Coulomb interactions arising from low dimensionality and reduced dielectric screening. Furthermore, as will be discussed in section II, inversion symmetry breaking and strong spin-orbit coupling in TMDCs lead to valley-selective circular dichroism[22,35,36,37,38]. Different valleys can be pumped selectively using circularly polarized light to generate carriers with different magnetic moments, leading to an emerging field – the "valleytronics". Hexagonal boron nitride (hBN) is another important type of 2D material[10,13]. It has a large bandgap of around 6 eV as shown in Figure 1b, making it an excellent dielectric. It can be incorporated in various heterostructures for electrostatic gating of other 2D materials, because "lattice match" is not necessary in these van der Waals heterostructures. Other than graphene, TMDCs, and boron nitride, the recently rediscovered black phosphorus (BP) shows a direct bandgap of around 0.3 eV in its bulk form[39]. The bandgap is expected to increase monotonically as the layer number decreases due to the electronic confinement in the direction perpendicular to the 2D plane, and its single layer is estimated to reach a single-particle bandgap of around 2 eV[40], as shown in Figure 1d. The wide range of material properties and the possibilities in combining 2D materials with different layer numbers and compositions allow for the realization of various nanophotonic devices and the exploration of fundamental optical sciences, covering a very wide spectral range from microwave to ultraviolet, as summarized in Figure 1a.

- *Figure 1: Two dimensional materials covering a broad spectral range*

## II. Photonic devices using 2D materials
**Photodetection using graphene**

Unique optical properties of 2D materials enable many important device applications in nanophotonics. Graphene attracts significant attentions for photodetection due to its strong interaction with photons in a wide energy range[33,34] and its high carrier mobility, making it a promising candidate for high speed applications in a broad wavelength range. In early works, the photodetection mechanism in graphene was attributed to the traditional photovoltaic (PV) effect[41,42,43] as in three-dimensional semiconductors such as gallium arsenide and silicon, in which the separation of photo-generated electron-hole pairs by built-in electric field leads to the photocurrent.

However, subsequent investigations pioneered by Xu *et al.* (ref. 44) revealed rather complex mechanisms for photocurrent generation in graphene, due to its gapless nature and reduced dimensionality. First, due to the acoustic phonon decay bottleneck and low heat capacity of single and few-layer graphene, the electronic temperature $T_E$ can be significantly higher than the lattice temperature $T_L$ under optical excitation[45,46]. Since the incidence light spot usually has a Gaussian-like intensity distribution, the resulting $T_E$ has a gradient with circular symmetry within the area of illumination. If the doping of graphene under illumination is uniform, there will be no net photocurrent generated due to $T_E$ gradient since the carriers will diffuse in all possible directions from the center to the edge within the illumination spot. However, if there is doping non-uniformity in graphene, for example in a graphene p-n junction, net photocurrent driven by these "hot" carriers can be observed, a phenomenon called photo-thermoelectric (PTE) effect[44]. Second, due to the zero gap nature, carrier-carrier scattering in graphene quickly (in sub pico-second) mixes photo-excited electrons and holes in conduction and valence bands through Auger-type processes, leading to a common quasi-Fermi level for both types of carriers[47,48,49]. At the same time, the Auger-type impact ionization due to strong carrier-carrier interaction can also lead to generation of multiple electron-hole pairs with a single photon, an effect called carrier multiplication[50,51]. The carrier multiplication in graphene occurs under zero external bias, fundamentally different from the multiplication effect in traditional avalanche photodetectors, in which carriers are first accelerated under a high electric field to acquire enough energy to excite additional electrons from valence to conduction bands. In short, the dynamics in graphene under illumination are different in many aspects from those in three-dimensional semiconductors with a significant bandgap, where the photo-generated electrons and holes are characterized by well-separated quasi-Fermi levels and the carrier temperature does not significantly differ from the lattice temperature.

Gabor *et al.* (ref. 52) demonstrated that at a dual gated graphene p-n junction as shown in Figure 2a, the photo-voltage exhibits a six-fold polarity variation pattern as the top and bottom gate biases change (Figure 2b), a characteristic of the PTE effect. The changing of photocurrent polarity for multiple times is attributed to the non-monotonic dependence of the graphene Seebeck coefficient on gate biases, which governs the photo-voltage (or current) due to the PTE effect. On the contrary, if the traditional PV effect dominates, the polarity of photocurrent only depends on the direction of the built-in electric field, and multiple polarity variations as a function of gate biases are not expected.

Freitag *et al.* (ref. 53) explored the photocurrent in biased graphene. Figure 2c denotes a reflection image of a graphene field-effect-transistor (FET) under investigation with a channel of 1 μm wide by 6 μm. A constant source-drain bias of 1 volt is applied to the device and photocurrent is determined by comparing the source-drain current with and without light illumination. In this case, three photocurrent generation mechanisms are identified: the traditional photovoltaic (PV), the photo-thermoelectric (PTE), and the bolometric effects. As shown in Figure 2d, when the channel doping is high (but uniform), the PTE effect is negligible as discussed above. The PV effect does exist but the dominant effect is the bolometric effect: the total current decreases under light excitation due to the reduction of the carrier mobility arising from the increase of the temperature. When the graphene channel is intrinsic (shown in Figure 2e, gate bias $V_G$ is close to Dirac point voltage $V_{Dirac}$), the traditional PV effect dominates, because the increase in carrier temperature leads to extra electrons and holes[54], leading to the enhancement of the total current. In intrinsic graphene, the carrier mobility is not sensitive to temperature[55] and bolometric effect can be ignored. Furthermore, although in this case finite drain bias $V_D$ will introduce a slightly non-uniform doping along the graphene channel, the PTE effect will reduce the total current under light illumination as shown in Figure 2e and from the polarity of the photocurrent, it is clear that in this case the PV effect dominates. Figure 2f summarizes the polarity of the photocurrent in biased graphene under different biasing conditions.

It has been shown experimentally by many groups that graphene photodetectors can operate at a high speed greater than tens of gigahertz[9,24,25,26,27]. Photocurrent due to the PV effect is intrinsically suitable for the high speed operations (up to hundreds of gigahertz) due to the high carrier mobility and carrier saturation velocity. For the PTE effect, the maximum speed of operation depends on the carrier cooling time, which can be longer than 100 pico-seconds[52]. This relatively long time in graphene can still support photodetectors operational at least in gigahertz range. However, the enhancement of photocurrent due to the PTE effect requires the suppression of cooling channels to increase the carrier temperature $T_E$, which will inevitably, lead to longer cooling time and lower speed. Effective utilization of multiple photocurrent generation mechanisms in graphene for different optical applications is an important future research topic. In addition, carrier multiplication effect may be leveraged to further enhance the performance of graphene photodetectors.

*- Figure 2: Photocurrent generation mechanisms in graphene*

**Light emitting properties and circular dichroism in TMDCs**

Unlike graphene, some members of the TMDC family are true semiconductors with appreciable band gaps; in particular, molybdenum- and tungsten-based dichalcogenides exhibit optical band gaps in the range of 1-2 eV, making them suitable for near-infrared absorption and emission. Interestingly, although these materials have an indirect gap in bulk and few-layer form, upon thinning down to monolayers they become direct-gap semiconductors with strong photoluminescence[14,15]. The band gaps in TMDCs have also been shown to be tunable - over a range from semiconducting to near metallic - via the application of external electric field as well as mechanical strain[56,57,58], useful for engineering the optoelectronic response for specific applications. More importantly, TMDC monolayers have been shown to display unique physics hitherto unobserved in other 2D materials. In contrast to conventional semiconductors such as gallium arsenide (GaAs), the direct gap in TMDC monolayers occurs at the two unequalled corners (K and K′) of the hexagonal Brillouin zone, endowing the electrons with a valley degree of freedom[22]. It was predicted that the lack of an inversion center of the crystal structure leads to valley-contrasting orbital magnetic moment and circular dichroism[35], which allows selective pumping of valley carriers by controlling the circular polarization of light[36,37,38]. It is also possible, vice versa, to make polarized LEDs in TMDC monolayer p-n junctions[59,60,61], as shown in Figure 3a. In this case, the electric field at the junction populates the carriers in a certain valley (K or K′) preferably, depending on the field direction. As a result, reversing the field direction switches the dominant polarization ($\sigma_+$ or $\sigma_-$) in the emitted light[59]. Such TMDC monolayer p-n junctions can also be used for light detection and energy harvesting[62]. Additionally, field-tunable valley magnetic moment and excitonic valley coherence have also been demonstrated[63]. This type of dynamic control of valley index opens up the exciting possibility of "valleytronics" - optoelectronic devices and systems based on the manipulation of the electrons' valley index. Furthermore, the strong spin-orbit coupling arising from transition metals locks the valley index with the electron spin, which could be useful to many optoelectronic applications. For example, the injection of carriers with specific spins, e.g., through ferromagnetic contacts, can create a population imbalance between the two valleys.

Another important property of TMDC monolayers is their giant exciton binding energy (~0.5-1 eV) arising from substantially reduced dielectric screening relative to the bulk[64,65]. This leads to strong and long-lived excitons, making them suitable for LEDs, photo-markers, etc. Figure 3b shows a reflection contrast spectrum of monolayer $WS_2$, where the Rydberg excitonic series are observed. Finally, the strong Coulomb interaction also makes possible the observation of trions, quasiparticles comprised of two electrons and a hole (or vice versa), in doped monolayer TMDCs under optical excitation[66,67]. Figure 3c denotes a photoluminescence spectrum of a monolayer $MoSe_2$ at 20 Kelvin. Both exciton ($X^0$) and trion ($X^-$) peaks are observed. From the energy spacing between $X^0$ and $X^-$, a trion bonding energy of 30 meV is determined. These quasiparticles, which are unstable in conventional semiconductors due to screened Coulomb interaction, have binding energies an order of magnitude larger than that of GaAs quantum wells and display similar valley-selective properties as the neutral excitons. However, unlike excitons, trions, being charged quasiparticles, are amenable to manipulation by electric fields, which could allow for exploration of charge transport of composite particles[68], and efficient collection of photo-generated current rather than relying on carrier diffusion alone.

- *Figure 3: Optoelectronic properties of single layer TMDCs*

**Energy harvesting and photodetection using 2D heterostructures**

Despite being atomically thin, the surfaces of 2D materials are self-passivated without any dangling bonds. As a result, it is possible to construct functional devices by combining different 2D materials with very different lattice constants to form heterostructures, leveraging the desirable property of each material. Britnell *et al.* (ref. 69) sandwiched a thin layer of tungsten disulfide ($WS_2$) within two layers of graphene and formed an efficient, "Schottky diode like" solar cell. In this device, the semiconducting $WS_2$ functions as the active energy harvesting material, while the metallic graphene works as a transparent electrode for efficient collection of photo-generated carriers. They achieved an extrinsic quantum efficiency of 30% in visible spectral range using merely around 50 nanometers of $WS_2$, revealing the great potential of TMDCs in solar energy harvesting. Although previously both $MoS_2$ and $WS_2$ have been explored for photovoltaic applications[70], the reintroduction of TMDCs in the unique 2D heterostructure form allows for greatly improved photo-carrier extraction efficiency. Photo-carriers only need to travel for less than tens of nanometers before being collected, providing a viable approach for the utilization of TMDCs in realistic photovoltaic applications. Yu *et al.* (ref. 71) reported a photodetector based on a graphene/MoS$_2$/graphene heterostructure and an external quantum efficiency of great than 50% is achieved at a wavelength of around 500 nm. In both heterostructures, the photoresponse can be tuned using a back gate underneath the bottom graphene contact, since a single layer of graphene cannot screen the electric field effectively, thus offering the flexibility in electrical tuning of the photoresponse.

Moreover, it is also possible to realize photocurrent gain in photodetection by integration of graphene and other photodetection materials. Konstantatos *et al.* (ref. 72) integrated lead sulfide (PbS) quantum dots with a graphene transistor and realized a phototransistor with a photo-conductive gain greater than $2\times10^8$. In this device, efficient light absorption is achieved using PbS quantum dots. Due to long carrier lifetime in PbS, photo-carriers can reach a high density, which effectively modify the graphene channel doping. Together with a high mobility in graphene, a large gain is achieved. Similar high gain photodetection was also observed in graphene/TMDC heterostructures[73]. In these devices, large gain usually implies slower device response. As a result, it is essential to optimize the gain and response time simultaneously for different applications.

## III. Enhancing the light-2D material interaction using photonic integration

Many two-dimensional materials interact with light strongly. The absorption coefficient in both graphene and single layer MoS$_2$ exceeds $5\times10^7$ m$^{-1}$ in visible if normalized to their atomic thickness[32], at least 10 times larger than those in gallium arsenide and silicon. However, due to their innate thinness, such strong interaction needs to be further enhanced for practical device applications. Integration of 2D materials with external photonic structures offers a solution. Moreover, integration with optical cavities allows for the manipulation of local optical density of states (DOS) significantly surrounding the 2D materials, leading to greatly modified emission/absorption properties.

Integration of ultrathin optical material and an optical waveguide is a classic approach to enhance the light-matter interaction in a non-resonant manner. Liu *et al.* (ref. 23) integrated wafer-scale monolayer graphene with a silicon optical waveguide and realized a broad band optical modulator covering the telecommunication wavelength from 1.3 to 1.6 μm, as shown in Figure 4a. Since a single layer of graphene absorbs around 2% of the vertical incidence light through inter-band transitions in near-infrared, it is only possible to modulate the transmission from 98% to 100% if graphene is used directly for light modulation purpose. By propagating the light in the silicon optical waveguide beneath the graphene as shown in Figure 4a, the light absorption is no longer limited to 2% but determined by the length of the waveguide. Using a 100 μm long silicon waveguide, a graphene modulator operational at gigahertz bandwidth with 10 dB modulation depth is demonstrated. Similar concept was adopted to enhance the responsivity of graphene high speed photodetectors for optical communications by three groups independently, as shown in Figures 4b to 4d[24,25,26]. A maximum responsivity of around 0.15 A/W and a 3dB bandwidth of around 20 GHz were realized in a broad wavelength range (from 1.3 to 1.6 μm), enabled by the non-resonant nature of the silicon optical waveguide. However, a relatively long waveguide length (from tens of microns to 100 microns) is necessary in order to achieve a high absorption, thus leading to a device length much longer that the wavelength and limiting the bandwidth of the photodetectors due to the large capacitance.

On the contrary, integration of 2D materials with an optical cavity makes the realization of compact devices possible at the expense of a reduced light-matter interaction bandwidth, which is determined by the quality factor of the cavity. Furchi *et al.* (ref. 27) integrated a graphene photodetector with a micro-cavity consisting of Bragg mirrors on both sides and achieved an enhancement of absorption of graphene by a factor of 26, as shown in Figure 4e. The limited enhancement is due to the relative low quality factor of the micro-cavity based on Bragg mirrors, because these mirrors only provide optical confinement in the vertical direction. Instead, the integration of graphene with high quality factor, silicon photonic crystal (PhC) cavity allows for much enhanced light-graphene interaction[29], because in such a cavity, confinement exists in all three-dimensions through in-plane Bragg reflections by periodic holes and total internal reflection in out-of-plane direction. Figure 4f shows such an integration scheme (top) and the silicon cavity used before (bottom left) and after (bottom right) the graphene deposition. Such a scheme can potentially be utilized to realize graphene modulators and detectors with ultracompact footprint and high modulation speed. However, the operation wavelength range will be limited by the cavity resonance wavelength and width. Furthermore, integration of cavities with single layer TMDCs allows for the modification of the spontaneous emission properties and may ultimately lead to TMDC based nanolasers[30,31].

- *Figure 4: Enhancing the light-2D material interaction using photonic integration*

## IV. Enhancing the light-2D material interaction using polaritonic resonances

Another approach to enhance the interaction of light and 2D materials is to use their intrinsic polaritonic resonances. Polaritons are quasiparticles resulting from the coupling of photons with an electric dipole-carrying elementary excitation such as plasmon, phonon, or exciton. The wide variety of 2D materials makes the exploration of different polaritonic excitations possible.

**Plasmon-polariton in graphene**

Propagating plasmon-polariton wave in metal-dielectric interface is routinely exploited to confine the light beyond the traditional "diffraction limit", leveraging the negative dielectric function in metal as shown in Figure 5a[74]. Because the dispersion relation (frequency $\omega$ vs. momentum $k$) of the light in free space is well above that of the plasmon-polariton wave, excitation of the propagating plasmon-polariton wave requires special $k$-matching techniques, in order to satisfy the boundary condition at the interface. Besides the propagating plasmon-polariton waves, highly confined, localized plasmon modes also exist in micro and nanostructures. In both cases, confinement beyond diffraction leads to highly concentrated light field at the interfaces and enhanced light-matter interaction. Direct excitation of the localized plasmon modes is possible because the $k$ needed is provided by the spatial confinement. Metallic graphene can support both propagating and localized plasmons[75,76,77,78,79,80].

Ju *et al.* (ref. 81) patterned a large piece of monolayer graphene into micro-ribbons as shown in Figure 5b and observed localized plasmon resonances in terahertz when the incident light polarization is perpendicular to the ribbons, because the ribbons can only be effectively polarized for plasmon excitation along the perpendicular direction. Figure 5c shows the extinction spectra for ribbons with widths of 1, 2, and 4 μm, indicating the momentum dependence of resonance since the momentum of localized plasmons is inversely proportional to the width. The plasmon resonance peak and amplitude can also be tuned using a gate bias, demonstrating the unique in-situ tunability. For graphene micro and nano-disks, localized plasmons can be excited regardless of the light polarization due to the high structural symmetry. Furthermore, utilization of multiple graphene layers effectively enhances the strength of the plasmon resonance through dipole-dipole couplings among layers, and infrared filters and polarizers were demonstrated using such graphene stacks[82]. Finally, the relativistic mass in graphene has been revealed in the exploration of plasmon resonances[81,82].

Another innovative approach to excited plasmons in graphene is demonstrated using a scanning near-field optical microscopy (SNOM) tip, as shown in the top panel of Figure 5d[83,84,85]. In this configuration, the nanoscale metallic tip is illuminated using the infrared light and the additional momentum required for the excitation of graphene plasmon is provided by the tip. The same tip is utilized to probe the plasmon waves in graphene. The bottom two panels in Figure 5d represent the measured and calculated light intensity at plasmonic resonance of 9.7 µm, respectively. Standing wave patterns are observed due to the reflections at the graphene edges.

Compared with plasmons in metals, graphene plasmon covers a relatively less explored wavelength range from terahertz to mid-infrared[86,87]. The electrical tunability makes the realization of optical modulators in this wavelength range possible. Together with the photo-thermoelectric effect discussed above, graphene plasmonic resonance may be utilized to construct high performance, tunable photodetectors and modulators covering a broad wavelength range from terahertz to mid-infrared. Furthermore, improvement of the graphene quality may lead to plasmon-polariton waveguide with high confinement beyond diffraction limit and propagation length exceeding tens of polariton wavelengths.

**Phonon-polaritons and plasmon-phonon polaritons**

As traditional polar substrates such as silicon oxide, 2D hexagonal boron nitride (hBN) supports phonon-polaritons, the quasiparticles due to the coupling of photons and dipole-carrying optical phonons. One major difference of phonon-polariton compared with plasmon-polariton is that transverse optical (TO) phonon can coupled to photon directly as long as the energy and momentum match, because light is also a transverse wave. In three-dimensional (3D) case, longitudinal optical (LO) phonons do not couple to light as plasmons. In fact, this is why in 3D plasmons cannot form polaritons. At the interface, the situation is different because the light incident at an angle can have a component parallel to the surface, which is a longitudinal wave. Dai *et al.* investigated the phonon-polaritons at the h-BN surface using the same SNOM technique for plasmon-polaritons, as shown in Figure 5e[88]. The layered nature of h-BN makes the tuning of phonon-polariton resonance in a layer-by-layer manner possible. The calculated dispersion relationships of phonon-polaritons at different h-BN thicknesses are shown in Figure 5f. The wide tuning range using hBN thickness may allow for the realization of mid-infrared optoelectronic devices in a broad wavelength range.

By placing the graphene on polar substrates such as silicon oxide or boron nitride, plasmons in graphene and phonons in polar substrates can couple if the energy and momentum match, leading to a new mixed state: plasmon-phonon polaritons[89,90]. Figure 5g illustrates such a coupling process and Figure 5h shows an extinction spectrum of a graphene/hBN ribbon array on silicon oxide substrate with a graphene ribbon width of around 300 nm. Multiple extinction peaks are due to the coupling of graphene plasmon, silicon oxide surface polar phonon, and hBN phonon modes. The exploration of the coupling of plasmons and phonons can not only provide useful information on the interaction of carriers and substrate, but also may further enable light detection and modulation functions, leveraging unique properties of both plasmons and phonons.
- *Figure 5: Enhancing the light-2D material using polaritonic resonances*

## V. Emerging layered black phosphorus for nanophotonics

Zero-gap graphene and single layer TMDCs such as $MoS_2$ and $WSe_2$ with a sizable optical bandgap greater than 1 eV are extensively explored, as discussed above. Recently rediscovered layered material black phosphorus (BP) with a direct bandgap of around 0.3 eV in its bulk form bridges the gap between the graphene and single layer TMDCs[39], making it an interesting addition to the existing 2D material family for nanophotonics and nanoelectronics[91,92,93,94,95,96]. Furthermore, by reducing the layer number, the bandgap of black phosphorus is expected to increase monotonically to around 2 eV (single-particle bandgap) in single layer, covering a broad energy range.

Xia *et al.* measured the infrared extinction spectra of thin-film (~ 30 nm) black phosphorus at different polarizations as shown in Figure 6a[94]. The extinction is directly proportional to optical conductivity and the strong polarization dependence indicates the anisotropic nature of the conductivity, arising from the puckered crystalline structure as shown in Figure 1a. Compared with graphene, the puckered black phosphorus has lower symmetry, which results in the in-plane anisotropic properties in momentum space[39,40,94,97]. Figure 6b plots the calculated thin-film BP conductivity along the x-direction at different doping levels for BP thicknesses of 4 and 20 nm, revealing the great potential of BP in light modulation functions[98]. The phononic properties of black phosphorus are also highly anisotropic and the Raman spectrum of thin-film black phosphorus (Figure 6c) also strongly depends on the polarization of the excitation light. In addition to the realization of mid- and near-infrared photonic devices using the thin-film black phosphorus, its unique anisotropic electronic, photonic, and phononic properties may allow for the realization of novel devices such as ballistic transistors and polarization sensors where the anisotropy is desirable.

It is also possible to construct photonic devices using heterostructures which consist of both narrow gap black phosphorus and larger gap TMDCs. Figure 6d denotes such an infrared light emitting diode. Here, the direct bandgap black phosphorus with a tunable gap by varying the layer number (0.3 eV is the minimal) is utilized as an active material for light emission. Injection of holes and electrons can be realized using p-type $WSe_2$ and n-type $MoS_2$, respectively. The injected carriers are trapped within black phosphorus for light emission due to the band-offset at the black phosphorus/TMDC interfaces. Graphene can be used to minimize the contact resistance[99]. If such a light emitting diode is integrated with a high-quality optical cavity, lasing is also possible. Phototransistors with a gain and electro-absorption modulators can also be built based on this concept of vertical van der Waals heterostructures.
- *Figure 6: Bridging the gap using black phosphorus*

## VI. Outlook

In summary, the emerging 2D materials provide the optical community with many exciting new opportunities for the exploration of sciences and technologies across a very wide electromagnetic spectral range. Graphene interacts with light strongly from terahertz to ultraviolet due to its gapless nature. Plasmon-polaritons in graphene allow for highly confined light field and greatly enhanced light-matter interaction at the resonance. Coupling of plasmon in graphene and phonon in polar dielectrics forms a new type of quasi-particle plasmon-phonon polariton, which can be utilized to enhance and tune the light-2D material interaction. Although graphene is not an optimal material for light emission, many single layer TMDCs are direct-bandgap semiconductors and exhibit strong excitonic emission properties that are gate-tunable, making them promising candidates for light emission in near-infrared. Furthermore, inversion-symmetry breaking accompanied by strong spin-orbit coupling in these single layers results in the unique valley polarization, leading to a new emerging field "valleytronics".

A recently rediscovered layered material, black phosphorus shows a bandgap of around 0.3 eV in its bulk form and its bandgap is expected to increase to around 2 eV in monolayer, thus bridging the gapless graphene and TMDCs with a sizable bandgap. The wide variety of 2D materials, together with approaches reviewed here (through photonic integration and through intrinsic polaritonic resonances) to enhance the light-matter interaction, may enable the discoveries of new optical sciences and the realization of various light emission, detection, modulation, and manipulation functions.

## Figure Captions

**Figure 1: Two dimensional materials covering a broad spectral range**
(a) A diagram of electromagnetic spectrum represented by a rainbow arrow. Applications utilizing the different spectral ranges are presented in the top portion of the panel. NIR, MIR, and FAR indicate near-, mid- and far-infrared, respectively. The atomic structures of hexagonal boron nitride (hBN), molybdenum disulfide ($MoS_2$), black phosphorus (BP), and graphene are shown in the bottom of the panel (from left to right). The crystalline directions (x and y) of anisotropic black phosphorus are indicated. The possible spectral ranges covered by different materials are indicated using colored polygons. (b), (c), (d), (e): bandstructures of single layer hBN, $MoS_2$, BP, and graphene, respectively.

**Figure 2: Photocurrent generation mechanisms in graphene**
(a) A schematic view of a dual-gated graphene field-effect-transistor for photocurrent generation experiment. (b) Main panel: six-fold photo-voltage pattern as the top and bottom gate biases change. The incident light spot is focused on the junction interface. Left (bottom): photo-voltage patterns taken along the vertical (diagonal) lines in the main panel. (a) and (b) are reprinted from ref. 52. (c) A reflection image of the graphene field-effect-transistor for biased photocurrent experiment. (d), (e) Energy diagrams and photocurrent generation mechanisms for strongly doped and intrinsic graphene channels, respectively. When the channel is strongly doped as shown in (d), the total photocurrent $I_{PC}$ is opposite to dc current $I_{DC}$, indicating the dominant component is bolometric photocurrent, $I_{BOL}$. On the contrary, in intrinsic graphene channel, the total photocurrent $I_{PC}$ is dominated by the traditional photo-voltaic component, $I_{PV}$. (f) Schematic diagram showing total photocurrent polarity with conditions (d) and (e) indicated by 1 and 2, respectively. "+" denotes photocurrent flowing from source to drain and "-" indicates opposite current flowing direction. (c) is reprinted from ref. 53. (d), (e), and (f) have been reproduced from their original version in ref. 53.

**Figure 3: Optoelectronic properties of TMDC monolayers**
(a) Polarization resolved electroluminescence spectra from a $WSe_2$ p-i-n junction. Left and right panels show the spectra dominated by σ- and σ+ circularly polarized light, respectively. The current flowing directions and K and K′ valleys are also shown. Reprinted from ref. 59. (b) The derivative of reflection contrast spectrum of $WS_2$ monolayer. Excitonic Rydberg series (A type excitons) are labelled by the quantum number like hydrogen series 1s, 2s, and so on. AX and $AX_T$ denote 1s exciton and trion transitions, respectively. Inset: reflection contrast spectrum. Main transition peaks correspond to A, B, and C excitons[67]. Reprinted from ref. 65. (c) Photoluminescence spectrum of monolayer $MoSe_2$ at 20 Kelvin. Inset: detailed view showing both exciton ($X^0$) and trion ($X^-$) peaks. The binding energy of $X^-$ is 30 meV. Reprinted from ref. 67.

**Figure 4: Enhancing the light-2D materials interaction using photonic integration**
(a) A schematic view of a high bandwidth graphene modulator integrated with a silicon waveguide. Reprinted from ref. 23. (b) Schematic of graphene/silicon heterostructure waveguide photodetector. Graphene/silicon heterojunctions are utilized for light detection. (c) A slightly different integration scheme of graphene photodetector and silicon waveguide. In this case, graphene functions as detection material and silicon waveguide confines and guides the light. (d)

A scanning electron micrograph of a graphene photodetector on silicon waveguide (false color). (b), (c), and (d) are reprinted from refs. 25, 24, and 26, respectively. (e) Photocurrent spectra measured from graphene photodetectors inside (red) and outside (blue) of an external Fabry-Perot cavity. Inset: schematic of the graphene detector inside the cavity. Reprinted from ref. 27. (f) Top panel: schematic of an electrically tunable graphene device integrated with a silicon nano-cavity. Lower left (right): scanning electron micrograph of the silicon cavity before (after) graphene deposition. Scale bar: 500 nm. Reprinted from ref. 29.

**Figure 5: Enhancing the light-2D material interaction using polaritonic resonances**
(a) Plasmon-polariton wave at the metal/dielectric interface. Due to the negative permittivity in metal, the field can be tightly confined at the interface. The field decaying length in both dielectric ($\delta_d$) and metal ($\delta_m$) can be much smaller than the wavelength of the light. (b) Atomic force micrographs of an array of graphene nanoribbons in which localized plasmons can be excited using vertical incident light with polarization perpendicular to the ribbons. (c) Relative extinction spectra due to localized plasmon resonance in ribbons with widths of 1, 2 and 4 µm, respectively. (b) and (c) are reprinted from ref. 81. (d) Top: the experimental schematic for the launch of the plasmon-polariton waves using a SNOM tip in a graphene wedge. The tip is illuminated using infrared light and the collection of local field in graphene is also realized using the tip. Middle (bottom): measured (calculated) the local field in graphene. Standing waves are formed due to the reflections at the edges. Reprinted from ref. 85. (e) The experimental schematic for the launch of the phonon-polariton waves using a SNOM tip in a hBN wedge. Inset at the lower left: schematic showing the relative motions of boron and nitride atoms. (f) Calculated phonon-polariton dispersion relations at different hBN thicknesses. (e) and (f) are reprinted from ref. 88. (g) Schematic showing the coupling of the graphene plasmon and hBN phonon. Reprinted from ref. 90. (h) Extinction in transmission in percentile for an array of graphene nanoribbons (300 nm wide) on a single layer of hBN on silicon oxide substrate. Peaks 1 and 2 are due to the coupling of graphene plasmon and $SiO_2$ phonon, while peaks 3 and 4 result from the coupling of graphene plasmon and hBN phonon.

**Figure 6: Bridging the gap using black phosphorus**
(a) Polarization resolved infrared extinction spectra of thin-film black phosphorus with a thickness of around 30 nm. Inset: an optical image of the BP thin film. The light polarizations used in measurements are shown in the image. Scale bar: 20 µm. The direction of zero degree is chosen randomly in the measurement and in this case is around 8 degrees off the x-axis of the BP crystal. Reprinted from ref. 94. (b) Calculated dynamical conductivities along x-direction of BP thin film at different doping levels. Red (grey) lines denote BP thickness of 20 (4) nm. Reprinted from ref. 98. (c) Polarization-resolved Raman scattering spectra of BP thin film under linearly polarized light excitation at 532 nm. D direction denotes light polarization 45 degrees to both x- and y- directions. Reprinted from ref. 94. (d) Schematic of a light emitting diode using 2D heterostructures. Narrow gap BP thin film is sandwiched between large gap TMDCs with p- and n-doping for the injection of holes and electrons, respectively.

# Figure 1: Two dimensional materials covering a broad spectral range

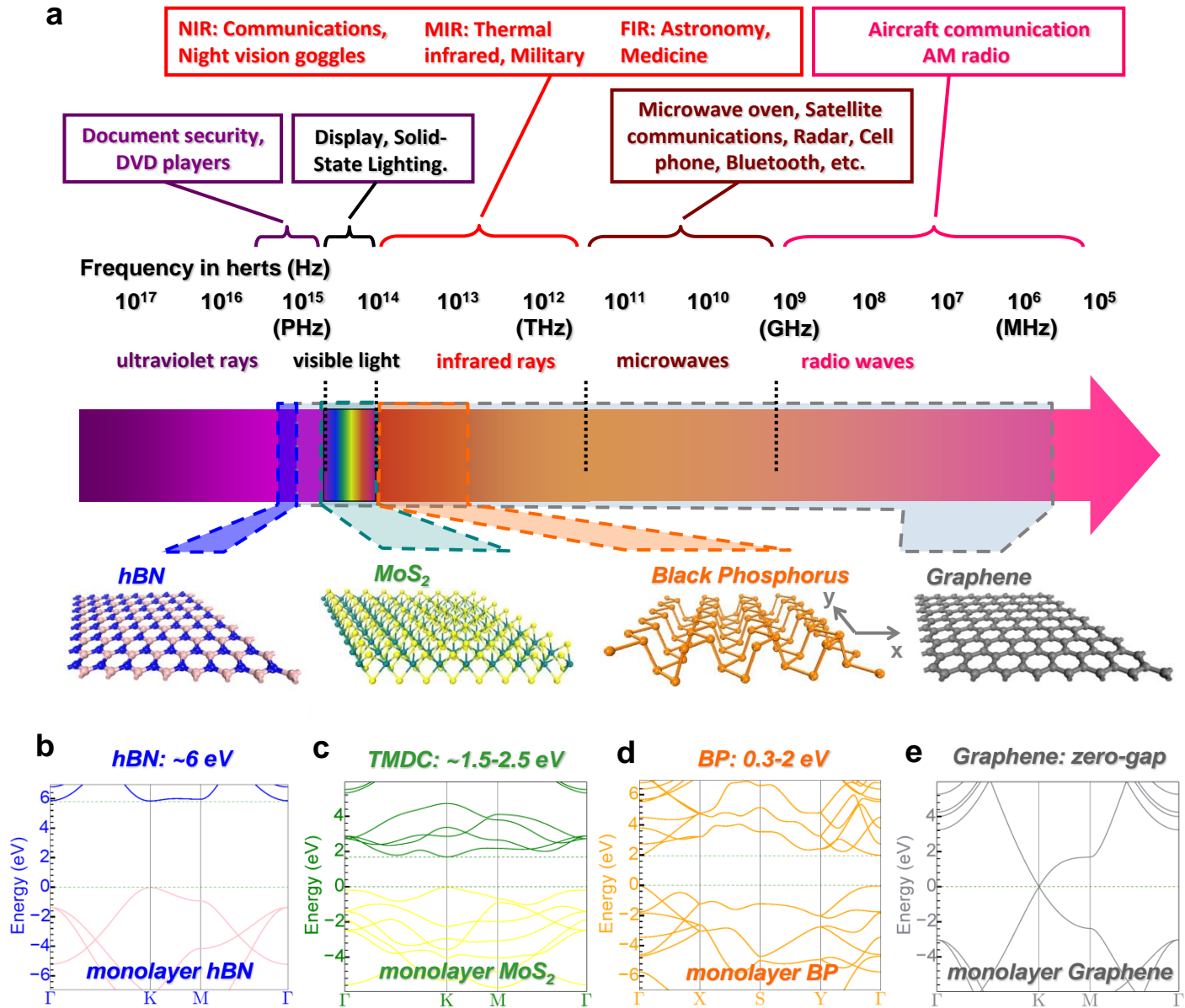

# Figure 2: Photocurrent generation mechanisms in graphene

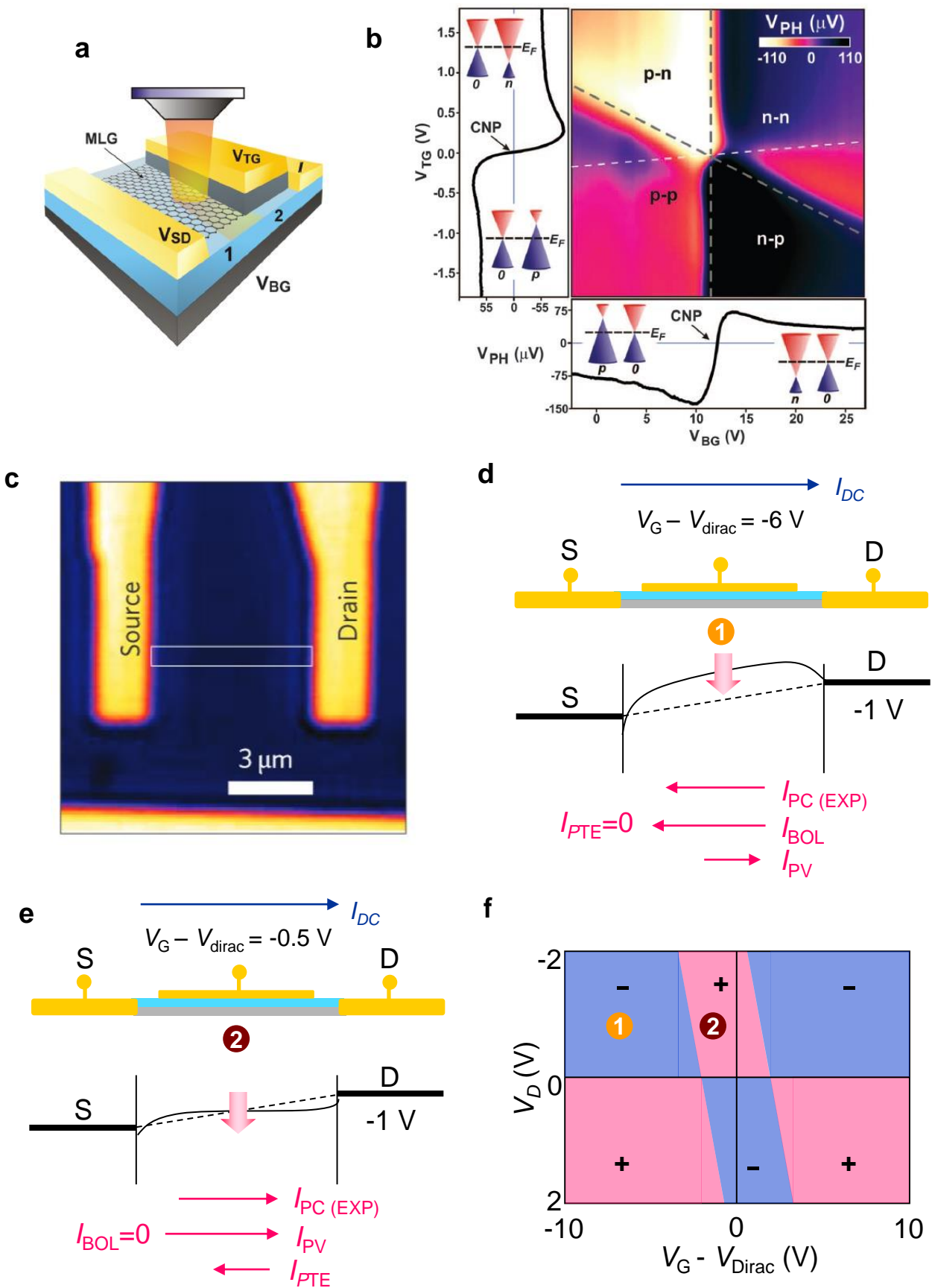

# Figure 3: Optoelectronic properties of TMDC monolayers

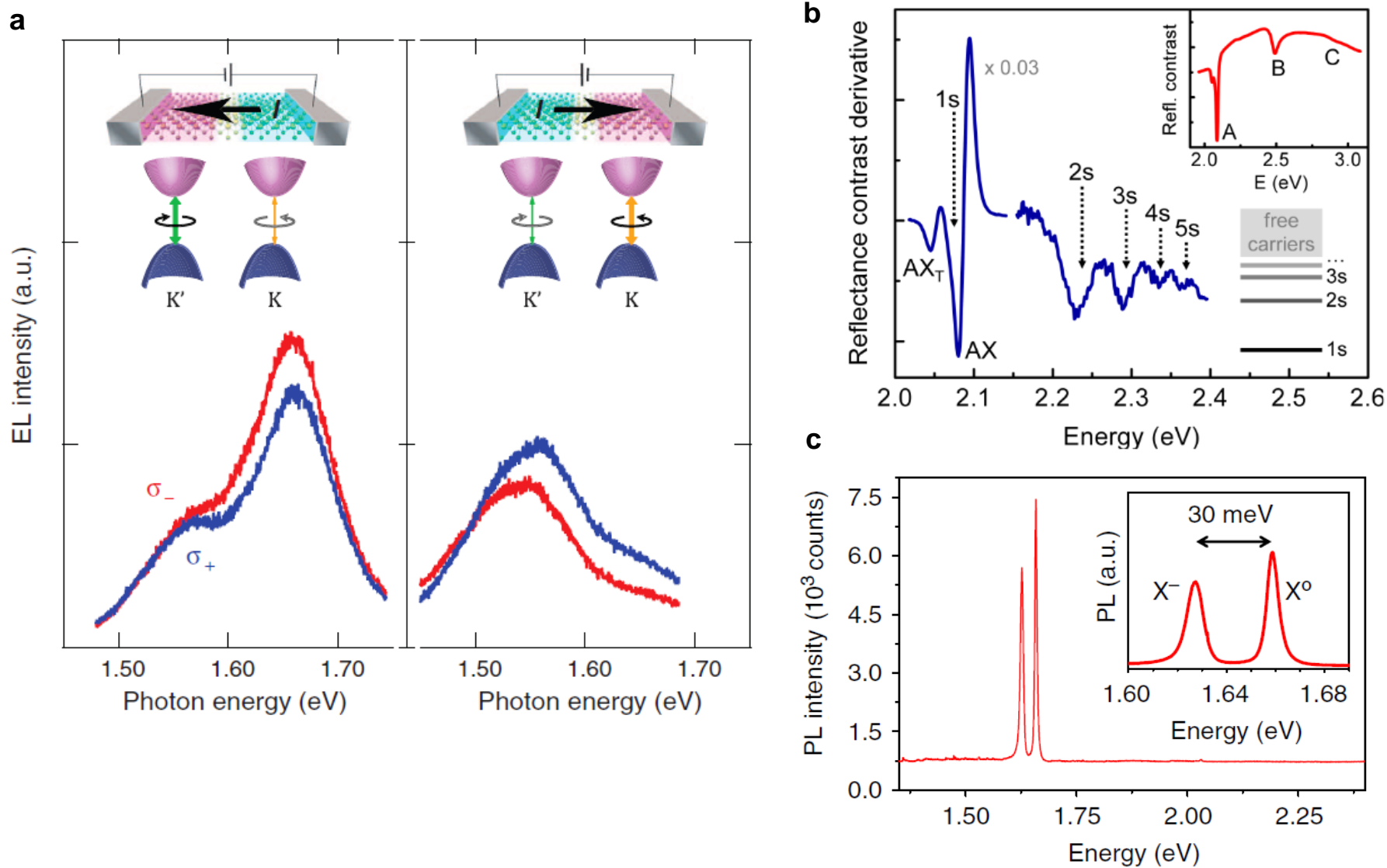

# Figure 4: Enhancing light-2D material interaction by photonic integration

## Waveguide Integration (non-resonant)

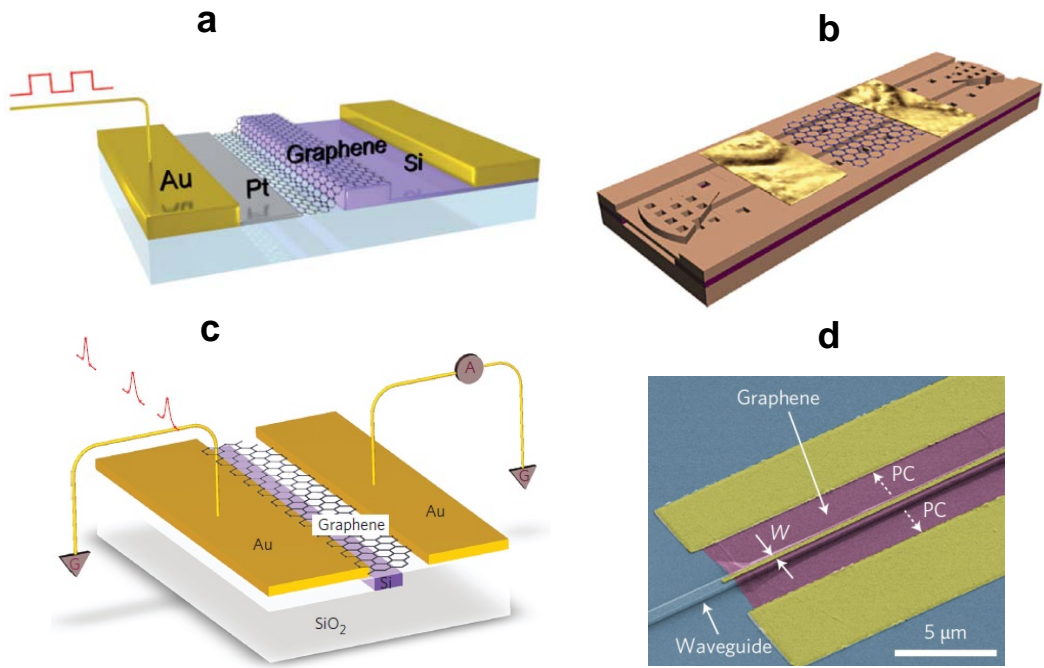

## Cavity Integration (resonant)

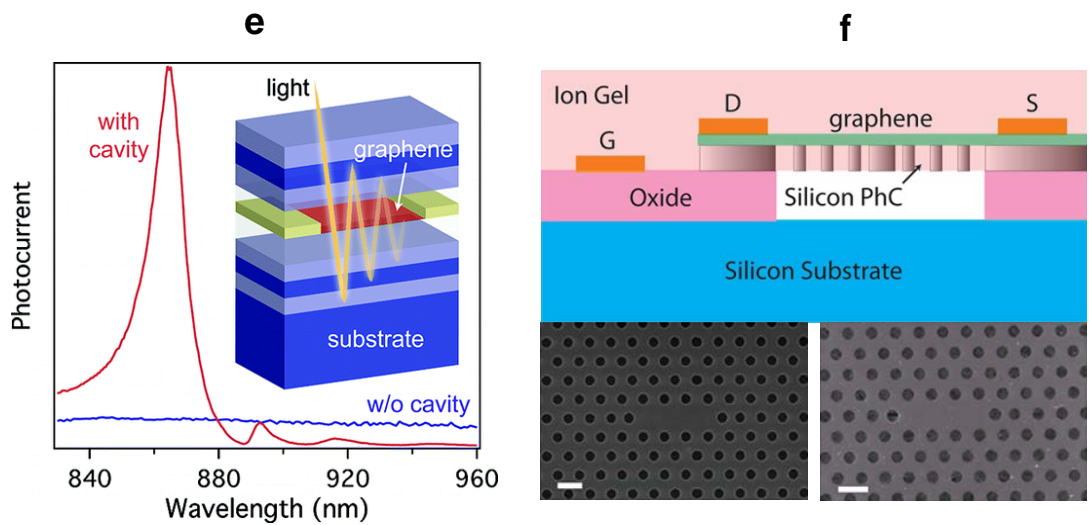

# Figure 5: Enhancing light-2D material interaction by polaritonic resonances

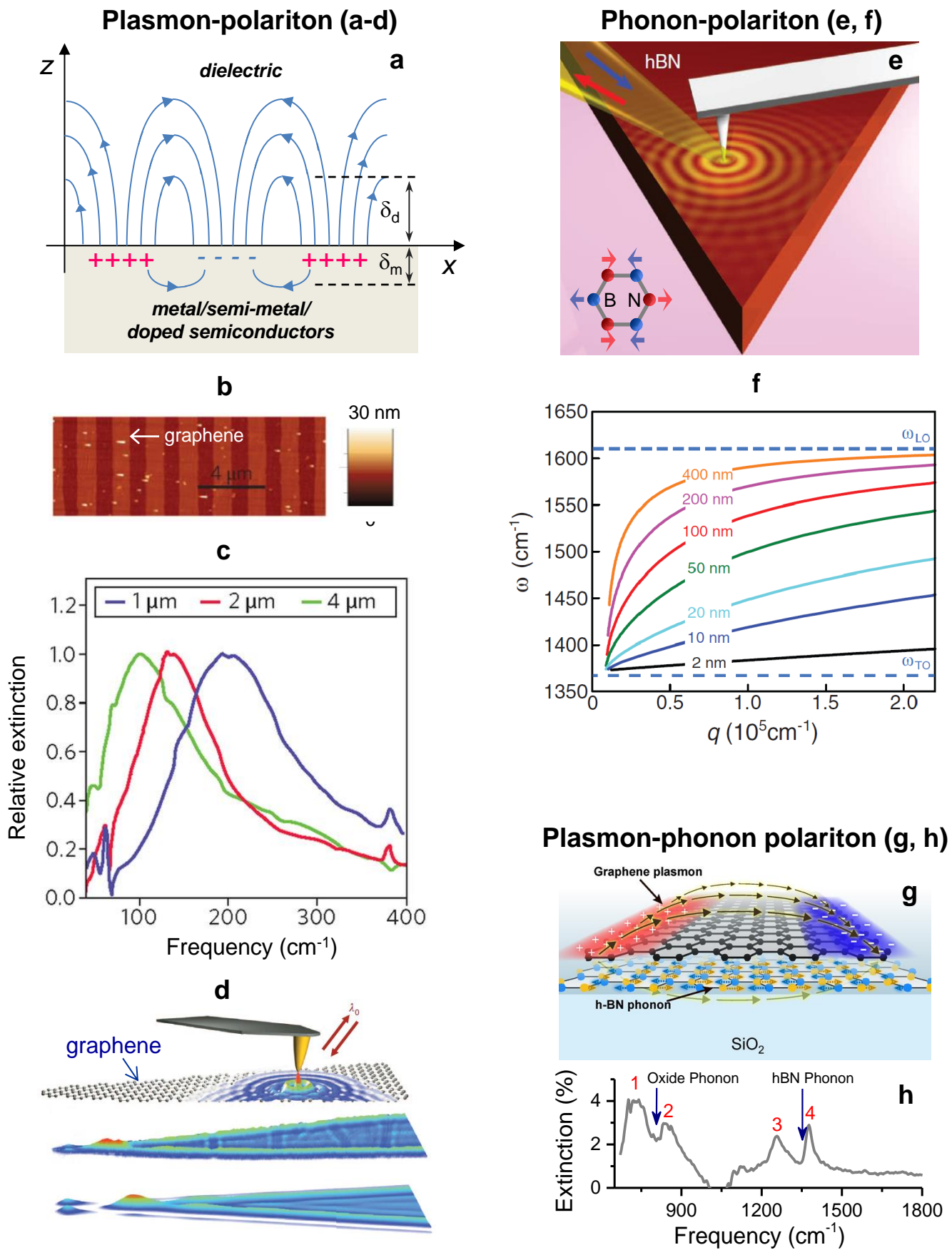

# Figure 6: Bridging the gap using black phosphorus

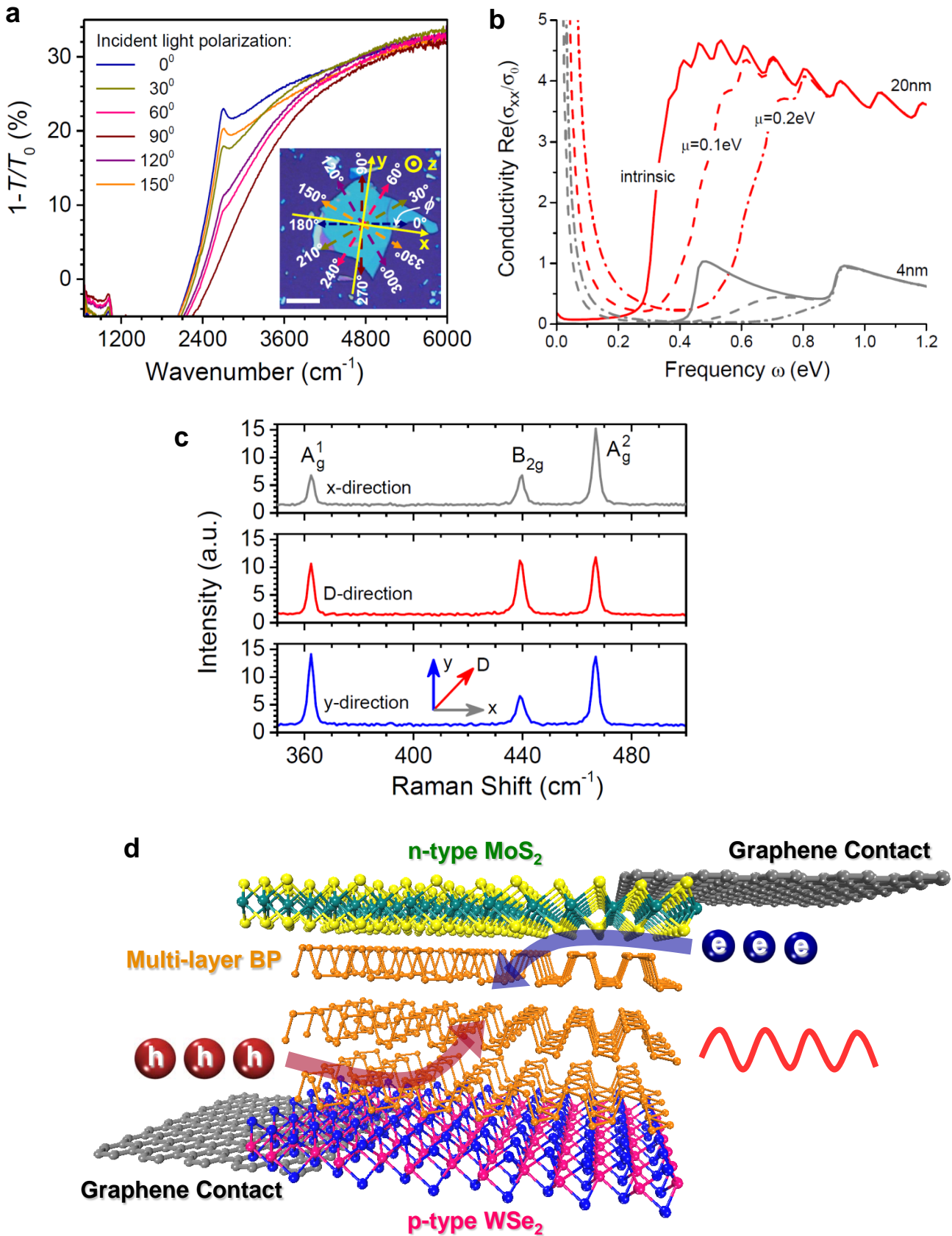